\documentclass[graybox]{svmult}

\usepackage{mathptmx}       % selects Times Roman as basic font
\usepackage{helvet}         % selects Helvetica as sans-serif font
\usepackage{courier}        % selects Courier as typewriter font
\usepackage{type1cm}        % activate if the above 3 fonts are
\usepackage{makeidx}         % allows index generation
\usepackage{graphicx}        % standard LaTeX graphics tool
\usepackage{multicol}        % used for the two-column index
\usepackage[bottom]{footmisc}% places footnotes at page bottom

\makeindex             % used for the subject index
\begin{document}

\newcommand\be{\begin{equation}}
\newcommand\ee{\end{equation}}
\newcommand\bea{\begin{eqnarray}}
\newcommand\eea{\end{eqnarray}}
\newcommand\tr{{\rm tr}\, }
\newcommand\nn{\nonumber \\}
\newcommand\e{\mathrm{e}}

\title*{Towards the unification of late-time acceleration and inflation by 
k-essence model %\footnote{Based on the work with J. Matsumoto \cite{Matsumoto:2010uv}}
}
\author{Shin'ichi Nojiri}
\institute{Shin'ichi Nojiri 
\at 
Department of Physics, Nagoya University, Nagoya 464-8602, Japan \\
and Kobayashi-Maskawa Institute for the Origin of Particles 
and the Universe,
%\footnote{See, http://www.kmi.nagoya-u.ac.jp/index-e.html .}
Nagoya University, Nagoya 464-8602, Japan 
\email{nojiri@phys.nagoya-u.ac.jp}}
\maketitle

\abstract*{
Based on the formulation of the reconstruction for the k-essence model, which was recently proposed in 
\cite{Matsumoto:2010uv}, we explicitly construct cosmological model to unifying the late-time 
acceleration and the inflation in the early universe. }

\abstract{Based on the formulation of the reconstruction for the k-essence model, which was recently proposed in 
\cite{Matsumoto:2010uv}, we explicitly construct cosmological model to unifying the late-time 
acceleration and the inflation in the early universe. }

\section{Introduction}

By several observations of the universe, it is widely believed that the expansion of 
the present universe is accelerating \cite{WMAP1,Komatsu:2008hk,SN1}. 
In order to explain the acceleration, many kinds of models have been proposed. 
In this report, we focus on so-called k-essence model 
\cite{Chiba:1999ka} in these models. 
The k-essence model is derived 
from k-inflation model \cite{ArmendarizPicon:1999rj} and
we may regard the tachyon dark energy 
model \cite{Sen:2002nu}, 
ghost condensation model \cite{ArkaniHamed:2003uy}, 
and scalar field quintessence model \cite{Peebles:1987ek}
with variations of the k-essence model. 

In this report, based on the formulation of the reconstruction \cite{Matsumoto:2010uv} 
for the k-essence model, we explicitly construct cosmological model to unifying the late-time 
acceleration and the inflation in the early universe 
(For consideration of unifying the inflation with the late-time acceleration 
in modified gravity, see \cite{Nojiri:2006ri}.). 
In \cite{Matsumoto:2010uv}, the k-essence models which reproduce the arbitrary 
FRW cosmology, that is, the arbitrary time-development of the scale factor or the Hubble rate, 
has been explicitly constructed. 
For general reconstruction, see \cite{Nojiri:2006be,reconstruction}.
In \cite{Matsumoto:2010uv}, two cases have been considered: One is the case that the action 
only contains the kinetic term and another is more general case including potential etc. 
In the former case, it was found that the exact $\Lambda$CDM model cannot be 
constructed although there is a model infinitely closing to $\Lambda$CDM model. 
In the model, however, the solution corresponding to $\Lambda$CDM model is 
unfortunately not stable. In the latter case, it has been found that there appear 
infinite number of redundant functions of the scalar fields, which are not directly 
related with the time development of the scale factor. 
By adjusting one of them, however, 
we can always obtain the model where the solution we need could be stable.

\section{Formulation of reconstruction}

In this section, we review on the formulation of the reconstruction proposed in 
\cite{Matsumoto:2010uv} and give the k-essence models which reproduce the arbitrary 
FRW cosmology, that is, the arbitrary time-development of the scale factor or the Hubble rate. 
We now consider a rather general model, whose action is given by
\be
\label{KK1}
S=  \int d^4 x \sqrt{-g} \left( \frac{R}{2\kappa^2} - K \left( 
\phi, X \right) + L_\mathrm{matter}\right)\, ,\quad X \equiv \partial^\mu \phi \partial_\mu \phi \, .
\ee
Here $\phi$ is a scalar field. 

Now the Einstein equation has the following form:
\be
\label{Sch2}
\frac{1}{\kappa^2}\left( R_{\mu\nu} - \frac{1}{2}g_{\mu\nu} R \right) 
= - K \left( \phi, X \right) g_{\mu\nu} + 2 K_X \left( \phi, X \right) \partial_\mu \phi \partial_\nu \phi 
+ T_{\mu\nu}\, .
\ee
Here $K_X \left( \phi, X \right) \equiv \partial K \left( \phi, X \right) / \partial X$ and 
$T_{\mu\nu}$ is the energy-momentum tensor of the matter. 
On the other hand, the variation of $\phi$ gives
\be
\label{Sch3}
0= - K_\phi \left( \phi, X \right) + 2 \nabla^\mu \left( K \left( \phi, X \right) \partial_\mu \phi \right)\, .
\ee
Here $K_\phi \left( \phi, X \right) \equiv \partial K \left( \phi, X \right) / \partial \phi$ and 
we have assumed that the scalar field $\phi$ does not directly couple with the matter. 

In this section, we assume the FRW universe whose spacial part is flat: 
$ds^2 = -dt^2 + a(t)^2 \sum _{i=1,2,3} (dx^i)^2$, 
and the scalar field $\phi$ only depends on time. 
Then the FRW equations are given by
\bea
\label{KK2}
&& \frac{3}{\kappa^2} H^2 = 2 X \frac{\partial K\left( \phi, X \right)}{\partial X} 
 - K\left( \phi, X \right) + \rho_\mathrm{matter}\, ,\nn
&& - \frac{1}{\kappa^2}\left(2 \dot H + 3 H^2 \right) 
= K\left( \phi, X \right) + p_\mathrm{matter}(t)\, .
\eea
We included the matters with constant EoS parameters $w_i$. 
Then the energy density of the matters is given by $\sum _i \rho_{0i}a^{-3(1+w_i)}$ with 
constants $\rho_{0i}$ and the pressure is given by $\sum _i w_i \rho_{0i}a^{-3(1+w_i)}$. 
Since the redefinition of $\phi$ can be absorbed into the redefinition of 
$K \left( \phi, X \right)$, we may identify the scalar field $\phi$ with the time 
coordinate $t$, $\phi=t$. 
Then we can rewrite the equations in (\ref{KK2}) in the following form
\bea
\label{KK4}
&& K\left( t, -1 \right) = - \frac{1}{\kappa^2}\left(2 \dot H + 3 H^2 \right) 
 - \sum_i w_i\rho_{0i} a^{-3(1+w_i)}\, ,\nn 
&& \left. \frac{\partial K\left( \phi, X \right)}{\partial X}\right|_{X=-1} 
=  \frac{1}{\kappa^2} \dot H  + \frac{1}{2}\sum_i \left(1+w_i\right) \rho_{0i} a^{-3(1+w_i)}\, .
\eea
By using the appropriate function $g(\phi)$, if we choose 
\bea
\label{KK5}
K(\phi,X) &=& \sum_{n=0}^\infty \left(X+1\right)^n K^{(n)} (\phi) \, , \nn
K^{(0)} (\phi) &\equiv & - \frac{1}{\kappa^2}\left(2 g''(\phi) + 3 g'(\phi)^2 \right)
 - \sum_i w_i\rho_{0i} a_0^{-3(1+w_i)}\e^{-3(1+w_i)g(\phi)} \nn 
K^{(1)} (\phi) &=& \frac{1}{\kappa^2} g''(\phi)  + \frac{1}{2}\sum_i \left(1+w_i\right) 
\rho_{0i} a_0^{-3(1+w_i)}\e^{-3(1+w_i)g(\phi)}\, ,
\eea
we find the following solution for the FRW equations (\ref{KK2}), 
\be
\label{KK6}
H= g'(t) \quad \left(a = a_0 \e^{g(t)} \right)\, .
\ee
Here $K^{(n)}(\phi)$ with $n=2,3,\cdots$ can be arbitrary functions. 
The case that $K^{(n)}(\phi)$'s with $n=2,3,\cdots$ vanish was studied in 
\cite{Nojiri:2005pu,Capozziello:2005tf} and the instability was investigated. 

It is often convenient to use redshift $z$ instead of cosmological time $t$ 
since the redshift has direct relation with observations. The redshift is defined by 
$a(t) = a\left(t_0\right)/\left(1+z\right) = \e^{N - N_0}$. 
Here $t_0$ is the cosmological time of the present universe, $N_0$ could be an arbitrary constant, 
and $N$ is called as e-folding and directly related with the redshift $z$. 
We now consider the reconstruction by using $N$ instead of the cosmological time $t$ and 
identify the scalar field $\phi$ with $N$
Then since $d/dt = H d/dN$, in terms of $N$, the equations in (\ref{KK4}) have 
the following expressions
\bea
\label{KK4B}
&& K\left( t, -H^2 \right) = - \frac{1}{\kappa^2}\left(2 H \frac{d H}{dN} + 3 H^2 \right) 
 - \sum_i w_i\rho_{0i} a^{-3(1+w_i)}\, ,\nn
&& \left. H^2 \frac{\partial K\left( \phi, X \right)}{\partial X}\right|_{X=-H^2} 
=  \frac{1}{\kappa^2} H \frac{d H}{dN} + \frac{1}{2}\sum_i \left(1+w_i\right) \rho_{0i} a^{-3(1+w_i)}\, .
\eea
By using the appropriate function $f(\phi)$, if we choose 
\bea
\label{KK5B}
K(\phi,X) &=& \sum_{n=0}^\infty \left(\frac{X}{f(\phi)^2} + 1\right)^n \tilde K^{(n)} (\phi) \, , \nn
\tilde K^{(0)} (\phi) &\equiv & - \frac{1}{\kappa^2}\left(2 f(\phi) f'(\phi) + 3 f(\phi)^2 \right)
 - \sum_i w_i\rho_{0i} a_0^{-3(1+w_i)}\e^{-3(1+w_i)\left( N - N_0 \right)} \nn 
\tilde K^{(1)} (\phi) &=& \frac{1}{\kappa^2} f(\phi) f'(\phi)  + \frac{1}{2}\sum_i \left(1+w_i\right) 
\rho_{0i} a_0^{-3(1+w_i)}\e^{-3(1+w_i) \left( N - N_0 \right) }\, ,
\eea
we find the following solution for the FRW equations (\ref{KK2}), 
\be
\label{KK6B}
H= f(N)\, , \quad \phi = N\, .
\ee
Note that if we define a new field $\varphi$ by 
$\varphi = \int \frac{d\phi}{f(\phi)}$, 
which can be solved as $\phi$ as a function of $\varphi$ as $\phi(\varphi)$ and 
we find $\varphi = t$ up to an additive constant corresponding to the constant of the integration. 
We can also identify the expansion in (\ref{KK5B}) with that in (\ref{KK5}): 
$\sum_{n=0}^\infty \left(\frac{X}{f(\phi)^2} + 1\right)^n \tilde K^{(n)} (\phi) 
= \sum_{n=0}^\infty \left(\tilde X + 1\right)^n K^{(n)} (\varphi)$. 
Here $\tilde X \equiv \partial^\mu \varphi \partial_\mu \varphi$, 
$K^{(n)} (\varphi) \equiv \tilde K^{(n)} \left(\phi\left(\varphi\right) \right)$. 
Then even for the expansion in (\ref{KK5B}), we can use the arguments about the stability 
and the existence of the Schwarzschild solution, which will be given in the following sections. 

\section{Stability of Solution }

We now investigate the stability of the solution (\ref{KK6}) or (\ref{KK6B}). 
First we consider the case without matter.
 From Eqs. (\ref{KK2}), we can derive the following equation which does not contain 
the variable $g^{\prime \prime}$,
\be
\label{KK7}
3 \frac{1-y^2}{1+X}X= -\frac{\dot H}{H^2}+ \frac{\kappa ^2}{H^2} 
\sum ^{\infty}_{n=2} \left( (n-1)X-n-1 \right)X(X+1)^{n-2}K^{(n)}(\phi)\, ,
\ee
where $y=\frac{g^{\prime}}{H}$. 
Using (\ref{KK7}), we can rewrite $dy/dN = \left(1/H\right)dy/dt$ 
in the form which does not contain $g$:
\bea
\label{KK8}
&& \frac{dy}{dN} = 3X \frac{1-y^2}{1+X} \left(\frac{\dot \phi}{X} +y \right) \nn
&& - \frac{\kappa^2}{H^2} \sum ^{\infty}_{n=2} 
\left[ ( \dot \phi +yX) \left( (n-1)X -n-1 \right) 
+ \dot \phi n(X+1) \right](X+1)^{n-2}K^{(n)}(\phi)\, .
\eea
When we consider the perturbation from a solution $\phi=t$ 
by putting $\phi = t + \delta \phi$ in (\ref{KK8}), we obtain
\be
\frac{d \delta \dot \phi}{dN} = \left[ -3- \frac{g^{\prime \prime}}{g^{\prime 2}} - \frac{d}{dN} 
\left\{ \frac{ \kappa ^2}{6g^{\prime 2}}(8K^{(2)}- \frac{2}{\kappa ^2}g^{\prime \prime}) \right\} \right] 
\delta \dot \phi\, .
\ee
If the quantity inside $[\ ]$ is negative, the fluctuation $\delta\dot\phi$ becomes exponentially smaller 
with time and therefore the solution becomes stable. 
Note that the stability is determined only in terms of $K^{(2)}$ and does not depend on other $K^{(n)}$ ($n\neq 2$). 
Then if we choose $K^{(2)}$ properly, the solution corresponding to arbitrary development of the universe becomes 
stable. 

We now investigate the stability when we include the matter. Then the equation corresponding to Eq.(\ref{KK7}) 
has the following form:
\bea
\label{KK18}
&& 3 \frac{1-y^2}{1+X}X = -\frac{\dot H}{H^2} + \frac{\kappa ^2}{H^2} \sum ^{\infty}_{n=2} 
\left( (n-1)X-n-1\bigg)X(X+1)^{n-2}K^{(n)}(\phi) \right. \\ 
&& + \frac{\kappa^2}{H^2} \frac{X-1}{2(X+1)} \rho_\mathrm{matter} 
 - \frac{\kappa^2}{2H^2}p_\mathrm{matter}-\frac{\kappa^2}{H^2}\frac{X}{X+1} 
\sum_i \rho _{0i} a_0 ^{-3(1+w_i)} \e^{-3(1+w_i)g(\phi)}\, . \nonumber
\eea
Then we find
\bea
\label{KK19}
\frac{dy}{dN} &=& 3X \frac{1-y^2}{1+X} \left(\frac{\dot \phi}{X} +y \right) \nn
&& - \frac{\kappa^2}{H^2} \sum^{\infty}_{n=2} \left[ ( \dot \phi +yX) \left( (n-1)X -n-1 \right) 
+ \dot \phi n(X+1) \right](X+1)^{n-2}K^{(n)}(\phi) \nonumber \\
&& + \frac{\kappa ^2}{2H^2 X} \left( -\frac{X-1}{X+1}\left( \dot \phi + yX \right) - \dot \phi \right) 
\rho_\mathrm{matter} + \frac{\kappa ^2}{2H^2}yp_\mathrm{matter} \nonumber \\
&& + \frac{\kappa ^2}{2H^2} \sum_i \left( \left( \dot \phi +yX \right) \frac{2}{X+1} - \dot \phi (1+w_i) \right) 
\rho _{0i} a_0^{-3(1+w_i)} \e^{-3(1+w_i)g(\phi)}\, .
\eea
When we include the matter, the situation becomes a little bit complicated since not only $H$ but the scale factor 
$a$ appears in the equation. Then we need equation to describe the time development of $a$. 
By defining $\delta\lambda$ for convenience as
\be
\label{KK24}
\delta \lambda \equiv 3 \sum _i (1+w_i) \rho _{0i} a(t)^{-3(1+w_i)} \frac{\kappa ^2}{6 g^{\prime 2}} 
\left( \frac{\delta a}{a} -g ^{\prime} \delta \phi \right)\, .
\ee
we obtain the following equations,
\bea
\label{KK29}
&& \left. \frac{d}{dN} \left(
\begin{array}{c}
 \delta \dot \phi \\
 \delta \lambda
\end{array}
\right)
\right| _{\phi =t, H=g^{\prime}(t)} =
 \left(
\begin{array}{cc}
A & B \\
C & D
\end{array}
\right)
\left(
\begin{array}{c}
 \delta \dot \phi \\
 \delta \lambda
\end{array}
\right) \, \\
%\label{KK30}
&& A \equiv -3 + \frac{g ^{\prime \prime}}{g^{\prime 2}} + \frac{\kappa ^2}{2 g^{\prime 2}} 
\sum _i (1+w_i) \rho _{0i} a(t)^{-3(1+w_i)} \nn
&& \qquad - \frac{d}{dN} \ln \left\{ 8K^{(2)} - \frac{2}{\kappa ^2}g^{\prime \prime} - \sum _i (1+w_i) 
\rho _{0i} a(t)^{-3(1+w_i)} \right\} \, ,\nonumber \\
&& B \equiv 3 - \frac{24 K^{(2)}}{8K^{(2)} - \frac{2}{\kappa^2} g^{\prime \prime} 
 - \sum _i (1+w_i) \rho _{0i} a(t)^{-3(1+w_i)}} \, , \nonumber \\
&& C \equiv 3 \sum _i (1+w_i) \rho _{0i} a(t)^{-3(1+w_i)} 
\left( \frac{\kappa ^2}{6 g^{\prime 2}} \right)^2 \nn
&& \qquad \times \left\{ 8K^{(2)}- \frac{6g^{\prime 2}}{\kappa ^2} - \frac{2}{\kappa ^2}g^{\prime \prime}
 - \sum _i (1+w_i) \rho _{0i} a(t)^{-3(1+w_i)}  \right\} \, , \nonumber \\
&& D \equiv \frac{d}{dN} \ln \left\{ \frac{\kappa ^2}{2 g^{\prime 2}} 
\sum _i (1+w_i) \rho _{0i} a(t)^{-3(1+w_i)} \right\} %\nn
%&& \qquad
 - \frac{\kappa ^2}{2g^{\prime 2}} \sum _i (1+w_i) \rho _{0i} a(t)^{-3(1+w_i)}\, . \nonumber
\eea
Generally, 2$\times$2 matrix must have negative trace and positive determinant in order that 
two eigenvalues could be negative since the two eigenvalues are given by 
$\frac{1}{2}\{ \tr M \pm \sqrt{(\tr M)^2-4(\det M)} \}$. 
So we just need to calculate the determinant and the trace of the matrix for investigating the stability 
of the fixed point $\phi =t$, $H=g^{\prime}(t)$. 
Since the expressions in (\ref{KK29}) are very complicated, we consider the case that the solution is given 
by the behavior mimicing $\Lambda$CDM solution in the Einstein gravity 
and the matter contents are given in the present universe. 
Then we find
\bea
\label{KK31}
&& a(t) \sim A \sinh ^{\frac{2}{3}} [\alpha t]\, , \quad %\\
%\label{KK32}
g^{\prime}(t) \sim \frac{2}{3} \alpha \coth [\alpha t]\, , \\
\label{KK33}
&& \frac{\kappa ^2}{\alpha ^2} \sum _i (1+w_i) \rho _{0i} a(t)^{-3(1+w_i)} 
\sim \frac{2}{3} \frac{1}{\sinh ^2 [\alpha t]}\, , \\
\label{KK34}
&& \frac{\kappa ^2}{\alpha ^2} \sum _i w_i (1+w_i) \rho _{0i} a(t)^{-3(1+w_i)} 
\sim \frac{4}{9} \times 1.86 \times 10^{-4} \frac{1}{\sinh ^{\frac{8}{3}}[\alpha t]}\, , 
\eea  
where $A \equiv (\rho _{m0}/ \rho _{\Lambda})^{\frac{1}{3}}$ and 
$\alpha \equiv \kappa \sqrt{3 \rho _{\Lambda}} /2$. 
Note that in (\ref{KK31}) %, (\ref{KK32}), 
and (\ref{KK33}), we neglect the contribution from radiation, and 
in (\ref{KK34}), there only appears the contribution from radiation. 
Therefore the expressions in (\ref{KK31}) - (\ref{KK34}) 
could be valid at least when $t \geq 10^9$ years. 
By using (\ref{KK31}) - (\ref{KK34}), we find the following expressions 
of the determinant and trace of the matrix 
in Eq.(\ref{KK29}):
\bea
\label{KK35}
&& \tr \left(
\begin{array}{cc}
A & B \\
C & D
\end{array}
\right) \sim
 -6 + \frac{3}{2} \frac{1}{\cosh ^2 [\alpha t]} - \frac{8 \frac{\kappa ^2}{\alpha ^3}K^{(2) \prime}
 - \frac{4}{3} \frac{\cosh [\alpha t]}{\sinh ^3 [\alpha t]} }
{ \frac{2}{3} \coth [\alpha t] \Big( 8 \frac{\kappa ^2}{\alpha ^2}K^{(2)} 
+ \frac{2}{3} \frac{1}{\sinh ^2 [\alpha t]} \Big)}\, , \\
\label{KK36}
&& \det \left(
\begin{array}{cc}
A & B \\
C & D
\end{array}
\right) \sim 
\left\{ 8 \frac{\kappa ^2}{\alpha ^2}K^{(2)} + \frac{2}{3} \frac{1}{\sinh ^2 [\alpha t]}  \right\} ^{-1} \nn
&& \times \left[ \left( -27 \frac{\sinh [\alpha t]}{\cosh ^3 [\alpha t]} + 36 \tanh [\alpha t]  \right) 
\frac{\kappa ^2}{\alpha ^3}K^{(2) \prime} \right. \nonumber \\
&& + \left( 72-36 \frac{1}{\cosh ^2 [\alpha t]} -18 \frac{1}{\cosh ^4 [\alpha t]} 
 -12 \times 1.86 \times 10^{-4} \frac{1}{\sinh ^{\frac{8}{3}}[\alpha t]} \right) 
\frac{\kappa ^2}{\alpha ^2}K^{(2)} \nonumber \\
&& - 6 \times 1.86 \times 10^{-4} \frac{1}{\cosh ^2[\alpha t] \sinh ^{\frac{8}{3}}[\alpha t]}
 - \frac{3}{2} \frac{1}{\cosh ^4 [\alpha t] \sinh ^2 [\alpha t]} \nonumber \\
&& \left. + 3 \frac{1}{\sinh ^2 [\alpha t] \cosh ^2 [ \alpha t]} + 4 \times 1.86 \times 10^{-4} 
\frac{1}{\sinh ^{\frac{8}{3}}[\alpha t]}  \right]\, .
\eea 
Note that $1 < \cosh [ \alpha t] \leq 2$ and $0 < \sinh [ \alpha t] \leq 1.7$ in evolution of the universe, 
For example, if we consider the case that $ K^{(2)} $ is constant, then the trace of the matrix is always negative 
and the determinant is positive when $ K^{(2)} \geq 0 $ since $72 \frac{\kappa ^2}{\alpha ^2}K^{(2)}$ 
and $3/( \sinh ^2 [\alpha t] \cosh ^2 [\alpha t])$ are dominant terms in Eq.(\ref{KK36}). 
Therefore even if $K^{(2)}$ is constant, the fixed point solution mimicking $\Lambda$CDM solution in the Einstein gravity 
becomes stable as long as $K^{(2)} \geq 0$. 

\section{Schwarzschild solution}

We now consider the condition that there could be the Schwarzschild or Schwarzschild-(A)dS solution 
for the action (\ref{KK1}). 

We now assume $\phi$ is a constant: $\phi = \phi_0$, 
or the change of the value of $\phi$ is very small. 
Then the Einstein equation (\ref{Sch2}) and the equation (\ref{Sch3}) given by the variation of $\phi$ 
reduce to 
\bea
\label{Sch5}
&& \frac{1}{\kappa^2}\left( R_{\mu\nu} - \frac{1}{2}g_{\mu\nu} R \right) 
= - K \left( \phi_0, 0 \right) g_{\mu\nu} + T_{\mu\nu}\, , \\
\label{Sch6}
&& 0= K_\phi \left( \phi_0, 0 \right) \, .
\eea
When $T_{\mu\nu}=0$, if $K \left( \phi_0, 0 \right)$ does not vanish, a solution of (\ref{Sch5}) is given by 
Schwarzschild-(A)dS space-time. 
On the other hand, if $K \left( \phi, 0 \right)$ vanishes, the Schwarzschild solution, which is asymptotically 
the Minkowski space-time, is a solution. 
The equation (\ref{Sch6}) requires that $K_\phi \left( \phi, 0 \right)$ has an extremum in general or 
$K \left( \phi, 0 \right)$ can be a constant independent of the value of $\phi$. 
Especially if $K \left( \phi, 0 \right)$ vanishes identically, Eq.(\ref{Sch5}) gives 
the Schwarzschild solution. 

We now write $K \left( \phi, X \right)$ as in (\ref{KK5}), 
$K \left( \phi, X \right) = \sum_{n=0} \left( 1 + X \right)^n K_n (\phi)$. 
Then if $K \left( \phi, 0 \right)=0$, that is, if $K \left( \phi, 0 \right)$ vanishes 
independent of the value of $\phi$, we find
\be
\label{Sch8}
\sum_{n=0} K_n (\phi) = 0\, .
\ee
Especially we may choose 
\be
\label{Sch9}
K_3 (\phi) = - K_0 (\phi) - K_1 (\phi) - K_2 (\phi) \, .
\ee

Then if the condition (\ref{Sch8}) or (\ref{Sch9}) is satisfied, the Schwalzschild space-time is always a solution 
independent of the value of $\phi$ as long as $\phi$ is a constant. 
Then any point source of matter makes the Schwarzschild space-time which generates the Newton potential. 
Then the correction to the Newton law could not appear. 
Note that the value of the scalar field $\phi$ changes by the evolution of the universe but as long as 
the condition (\ref{Sch8}) or (\ref{Sch9}) is satisfied, in a local region where $\phi$ is almost constant, 
the correction to the Newton law could be negligible. 

\section{Construction of a model unifying late-time acceleration and inflation}

In this section, by using the formulation of the reconstruction, we try to construct a model describing the 
late-time acceleration and the inflation in the early universe. 

In the Einstein gravity, we have the following FRW equation
\be
\label{ku0}
\frac{3}{\kappa^2} H^2 = \rho_\mathrm{total}\, , 
\quad - \frac{1}{\kappa^2} \left(2\dot H + 3H^2 \right) = p_\mathrm{total}\, .
\ee
Here $\rho_\mathrm{total}$ and $p_\mathrm{total}$ are total energy density and the total pressure. 
Then we may define total equation of state parameter $w_\mathrm{total}$ by
\be
\label{ku00}
w_\mathrm{total} = \frac{p_\mathrm{total}}{\rho_\mathrm{total}} 
= -1 - \frac{2\dot H}{3H^2} = -1 - \frac{2}{3H} \frac{dH}{dN}\, .
\ee

As matters, we include the radiation and dust, which may corresponds to the baryon and dark matter. 
Then as in the energy density of matter behaves as in 
(\ref{KK5B}),
\be
\label{ku1}
\rho = \rho_\mathrm{radiation} a_0^{-4}\e^{-4 \left( N - N_0 \right) } 
+ \rho_\mathrm{dust} a_0^{-3}\e^{-3 \left( N - N_0 \right) }\, .
\ee
We may consider the following Hubble rate 
\be
\label{ku2}
H^2 = H_0^2 \left( N_I^2 + N^2 \right)^{-\gamma} 
+ \frac{\kappa^2}{3} \left( \rho_\mathrm{radiation} a_0^{-4}\e^{-4 \left( N - N_0 \right) } 
+ \rho_\mathrm{dust} a_0^{-3}\e^{-3 \left( N - N_0 \right) }\right) \, .
\ee
Here $H_0$, $N_I$, and $\gamma$ are parameters assumed to be positive. 
Then by using (\ref{KK4B}), we find
\bea
\label{ku3}
K^{(0)} (\phi) &=& \frac{H_0^2}{\kappa^2} \left\{2 \gamma \phi \left( N_I^2 + \phi^2 \right)^{-\gamma-1} 
+ 3 \left( N_I^2 + \phi^2 \right)^{-\gamma} \right\} \, , \nn
K^{(1)} (\phi) &=& - \frac{H_0^2}{\kappa^2} \gamma \phi \left( N_I^2 + \phi^2 \right)^{-\gamma-1}\, .   
\eea
Note that in the expression of $K^{(0)} (\phi)$ and $K^{(1)} (\phi)$, the parameters describing 
the matters like $\rho_\mathrm{radiation}$ and $\rho_\mathrm{dust}$ are not included. 

We now investigate the cosmology described by the Hubble rate in (\ref{ku2}). 
If $H_0$ is large enough, in the early universe, where we assume $N\to 0$, the first term in (\ref{ku2}) 
could dominate $H^2 \sim H_0^2 \left( N_I^2 + N^2 \right)^{-\gamma}$. Then by using (\ref{ku00}), we find
\be
\label{ku4}
w_\mathrm{total} \sim -1 + \frac{2\gamma N}{3\left( N_I^2 + N^2 \right)} \, .
\ee
Then $N\to 0$, we find $w_\mathrm{total} \to -1$, which corresponding to the effective cosmological 
constant with $w=-1$. Therefore the inflation in the early universe could be generated. 
When $N$ becomes larger, the first term could become smaller and the second term in (\ref{ku2}) corresponding 
to the radiation could dominate. Maybe more exactly, the second and third terms could be generated by 
the reheating after the inflation, after that there could be complicated process for the matters like 
pair annihilation (creation), baryogenesis (or leptogenesis). In this present report, we do not discuss 
about the detailed process. 
After the radiation, the third term in (\ref{ku2}) corresponding to the dust could dominate as in the usual 
scenario. The contributions from the radiation and matter decrease exponentially as a function of $N$, 
the first term in in (\ref{ku2}) dominate in the late universe again. Then $w_\mathrm{total}$ is given 
by (\ref{ku4}) again. For large $N$, corresponding to the late universe, we find $w_\mathrm{total}\to -1$, 
again and therefore the expansion of the universe accelerates again. 
Note that, however, different from the case of the small cosmological constant, $H$ is not a constant but 
decreasing function of $N$ as $H\sim H_0 N^{-\gamma}$. 
Then the smallness of the scale of the acceleration in the present universe might be naturally explained. 

We now discuss more quantitively. Since $H\to H_0 N_I^{-\gamma} $ when $N\to 0$, 
if we assume $N_I\sim \mathcal{O}(1)$, the scale of $H_0$ corresponds to the weak scale. 
In the action (\ref{KK1}) with (\ref{ku3}), the dimensional parameter appears only in the combination of
\be
\label{ku5}
M^4 = \frac{H_0^2}{\kappa^2} = M_\mathrm{Planck}^2 H_0^2\, .
\ee
Here $M_\mathrm{Planck}\sim 10^{19}\,\mathrm{GeV}$ is the Planck scale. Then if the scale 
of the inflation is the Planck scale, $H_0 \sim M_\mathrm{Planck}$, $M$ is also the Planck 
scale, $M \sim M_\mathrm{Planck}$. If $H_0$ is a GUT scale, $H_0 \sim 10^{16}\,\mathrm{GeV}$, 
the scale of $M$ is $10^{17\sim 18}\,\mathrm{GeV}$. 
We should note that $M$ is only one parameter with the dimension of mass in the action. 
We now consider how very small scale corresponding to the late acceleration can appear from only one 
dimensional parameter. 
In the present universe, if we denote the present value of $H$ by $H_\mathrm{present} 
\sim 10^{-33}\, \mathrm{eV}$, under the assumption $N_I\sim \mathcal{O}(1)$, 
we find $C \equiv \frac{\mbox{Plank scale}}{H_0} \sim 10^{61} \sim \e^{140}$. 
Then when we assume $H_0 \sim M_\mathrm{Planck}$, we have $\frac{H_0}{H_\mathrm{present}} \sim C$. 
If we naively assume $N \sim \ln C \gg N_I$, we find $\gamma \sim 28$. 
Then we find that the very small scale corresponding to the late acceleration can appear 
from only one dimensional parameter $M$. 
Then the fine tuning problem might be relaxed. 
If the equation of state parameter $w_\mathrm{dark energy}$ of the dark energy could be given 
by (\ref{ku4}) with $N\gg N_I$ as $w_\mathrm{dark energy}\sim w_\mathrm{total}$, we obtain 
$w_\mathrm{dark energy}\sim - 0.87$, 
which is little bit greater than the value obtained 
from the observation $-0.14 < 1+w < 0.12$\cite{Komatsu:2008hk}.

\section{Summary}

In this report, we explicitly construct cosmological model of k-essence to unifying the late-time 
acceleration and the inflation in the early universe 
(Note that such reconstruction scheme for k-essence models may be extended 
also for presence of Lagrange multiplier \cite{Capozziello:2010uv}.). 
The construction is based on the formulation of the reconstruction for the k-essence model, 
which was recently proposed in \cite{Matsumoto:2010uv}. 
The action (\ref{KK1}) can be expanded as a power series of 
$1 + X = 1 + \partial^\mu \phi \partial_\mu \phi$ as in (\ref{KK5}). 
For the cosmological evolution of the Hubble rate $H$, only the first two terms in the series are 
relevant. The third term is relevant for the stability. The fourth or higher terms are relevant for 
the existence of the Schwarzschild solution, which may reproduce the Newton law if the scalar field 
is not directly coupled with the matter. 

In the constructed model (\ref{ku3}), only one dimensional parameter, whose scale could be equal to 
or a little bit smaller than the Planck scale but the small scale of the present Hubble scale can be 
produced. 
We have not discussed about the reheating and structure formation in the universe etc., which will be 
discussed in the forthcoming paper.

\section*{Acknowledgments}

We are grateful to J.~Matsumoto and S.~D.~Odintsov for very helpful discussions.
This work has been supported by Global COE Program of Nagoya
University (G07) provided by the Ministry of Education,Culture, Sports,
Science
\& Technology of Japan, and by the JSPS Grant-in-Aid for Scientific Research
(S) \# 22224003.

%%%%%%%%%%%%%%%%%%%%%%%%%%%%%%%%%%%%%%


\begin{thebibliography}{99}

%\cite{Matsumoto:2010uv}
\bibitem{Matsumoto:2010uv}
  J.~Matsumoto and S.~Nojiri,
  %``Reconstruction of k-essence model,''
  Phys.\ Lett.\  B {\bf 687}, 236 (2010)
  [arXiv:1001.0220 [hep-th]].
  %%CITATION = PHLTA,B687,236;%%

\bibitem{WMAP1}
D.~N.~Spergel {\it et al.} [WMAP Collaboration],
Astrophys.\ J.\ Suppl.\  {\bf 148}, 175 (2003);\
H.~V.~Peiris {\it et al.}  [WMAP Collaboration],
\textit{ibid}. {\bf 148}, 213 (2003);\
D.~N.~Spergel {\it et al.}  [WMAP Collaboration],
\textit{ibid}. {\bf 170}, 377 (2007).

\bibitem{Komatsu:2008hk}
E.~Komatsu {\it et al.}  [WMAP Collaboration],
Astrophys.\ J.\ Suppl.\  {\bf 180}, 330 (2009).

\bibitem{SN1}
S.~Perlmutter {\it et al.}  [Supernova Cosmology Project Collaboration],
Astrophys.\ J.\  {\bf 517}, 565 (1999);\
A.~G.~Riess {\it et al.}  [Supernova Search Team Collaboration],
Astron.\ J.\  {\bf 116}, 1009 (1998);\
P.~Astier {\it et al.}  [The SNLS Collaboration],
Astron.\ Astrophys.\  {\bf 447}, 31 (2006);\
A.~G.~Riess {\it et al.},
Astrophys.\ J.\  {\bf 659}, 98 (2007).

%%% k-essence

%\cite{Chiba:1999ka}
\bibitem{Chiba:1999ka}
  T.~Chiba, T.~Okabe and M.~Yamaguchi,
  %``Kinetically driven quintessence,''
  Phys.\ Rev.\  D {\bf 62}, 023511 (2000)
  [arXiv:astro-ph/9912463];\ 
  %%CITATION = PHRVA,D62,023511;%%
%\cite{ArmendarizPicon:2000dh}
%\bibitem{ArmendarizPicon:2000dh}
  C.~Armendariz-Picon, V.~F.~Mukhanov and P.~J.~Steinhardt,
  %``A dynamical solution to the problem of a small cosmological constant  and
  %late-time cosmic acceleration,''
  Phys.\ Rev.\ Lett.\  {\bf 85}, 4438 (2000)
  [arXiv:astro-ph/0004134];\  
  %%CITATION = PRLTA,85,4438;%%
%\cite{ArmendarizPicon:2000ah}
%\bibitem{ArmendarizPicon:2000ah}
  C.~Armendariz-Picon, V.~F.~Mukhanov and P.~J.~Steinhardt,
  %``Essentials of k-essence,''
  Phys.\ Rev.\  D {\bf 63}, 103510 (2001)
  [arXiv:astro-ph/0006373].
  %%CITATION = PHRVA,D63,103510;%%

%%% k-inflation

%\cite{ArmendarizPicon:1999rj}
\bibitem{ArmendarizPicon:1999rj}
  C.~Armendariz-Picon, T.~Damour and V.~F.~Mukhanov,
  %``k-Inflation,''
  Phys.\ Lett.\  B {\bf 458}, 209 (1999)
  [arXiv:hep-th/9904075];\ 
  %%CITATION = PHLTA,B458,209;%%
%\cite{Garriga:1999vw}
%\bibitem{Garriga:1999vw}
  J.~Garriga and V.~F.~Mukhanov,
  %``Perturbations in k-inflation,''
  Phys.\ Lett.\  B {\bf 458}, 219 (1999)
  [arXiv:hep-th/9904176].
  %%CITATION = PHLTA,B458,219;%%

%%% tachyon dark energy

%\cite{Sen:2002nu}
\bibitem{Sen:2002nu}
  A.~Sen,
  %``Rolling Tachyon,''
  JHEP {\bf 0204}, 048 (2002)
  [arXiv:hep-th/0203211];\ 
  %%CITATION = JHEPA,0204,048;%%
%\cite{Sen:2002an}
%\bibitem{Sen:2002an}
  A.~Sen,
  %``Field theory of tachyon matter,''
  Mod.\ Phys.\ Lett.\  A {\bf 17}, 1797 (2002)
  [arXiv:hep-th/0204143];\ 
  %%CITATION = MPLAE,A17,1797;%%
%\cite{Gibbons:2002md}
%\bibitem{Gibbons:2002md}
  G.~W.~Gibbons,
  %``Cosmological evolution of the rolling tachyon,''
  Phys.\ Lett.\  B {\bf 537}, 1 (2002)
  [arXiv:hep-th/0204008];\ 
  %%CITATION = PHLTA,B537,1;%%
%\cite{Bagla:2002yn}
%\bibitem{Bagla:2002yn}
  J.~S.~Bagla, H.~K.~Jassal and T.~Padmanabhan,
  %``Cosmology with tachyon field as dark energy,''
  Phys.\ Rev.\  D {\bf 67}, 063504 (2003)
  [arXiv:astro-ph/0212198].
  %%CITATION = PHRVA,D67,063504;%%

%%% ghost condensation

%\cite{ArkaniHamed:2003uy}
\bibitem{ArkaniHamed:2003uy}
  N.~Arkani-Hamed, H.~C.~Cheng, M.~A.~Luty and S.~Mukohyama,
  %``Ghost condensation and a consistent infrared modification of gravity,''
  JHEP {\bf 0405}, 074 (2004)
  [arXiv:hep-th/0312099];\ 
  %%CITATION = JHEPA,0405,074;%%
%\cite{ArkaniHamed:2003uz}
%\bibitem{ArkaniHamed:2003uz}
  N.~Arkani-Hamed, P.~Creminelli, S.~Mukohyama and M.~Zaldarriaga,
  %``Ghost Inflation,''
  JCAP {\bf 0404}, 001 (2004)
  [arXiv:hep-th/0312100].
  %%CITATION = JCAPA,0404,001;%%

%%% quintessence

%\cite{Peebles:1987ek}
\bibitem{Peebles:1987ek}
  P.~J.~E.~Peebles and B.~Ratra,
  %``Cosmology with a Time Variable Cosmological Constant,''
  Astrophys.\ J.\  {\bf 325}, L17 (1988);\ 
  %%CITATION = ASJOA,325,L17;%%
%\cite{Ratra:1987rm}
%\bibitem{Ratra:1987rm}
  B.~Ratra and P.~J.~E.~Peebles,
  %``Cosmological Consequences of a Rolling Homogeneous Scalar Field,''
  Phys.\ Rev.\  D {\bf 37}, 3406 (1988);\ 
  %%CITATION = PHRVA,D37,3406;%%
%\cite{Chiba:1997ej}
%\bibitem{Chiba:1997ej}
  T.~Chiba, N.~Sugiyama and T.~Nakamura,
  %``Cosmology with x-matter,''
  Mon.\ Not.\ Roy.\ Astron.\ Soc.\  {\bf 289}, L5 (1997)
  [arXiv:astro-ph/9704199];\ 
  %%CITATION = MNRAA,289,L5;%%
%\cite{Zlatev:1998tr}
%\bibitem{Zlatev:1998tr}
  I.~Zlatev, L.~M.~Wang and P.~J.~Steinhardt,
  %``Quintessence, Cosmic Coincidence, and the Cosmological Constant,''
  Phys.\ Rev.\ Lett.\  {\bf 82}, 896 (1999)
  [arXiv:astro-ph/9807002].
  %%CITATION = PRLTA,82,896;%%

%\cite{Nojiri:2006ri}
\bibitem{Nojiri:2006ri}
  S.~Nojiri and S.~D.~Odintsov,
  %``Introduction to modified gravity and gravitational alternative for dark
  %energy,''
  eConf {\bf C0602061}, 06 (2006)
  [Int.\ J.\ Geom.\ Meth.\ Mod.\ Phys.\  {\bf 4}, 115 (2007)]
  [arXiv:hep-th/0601213].
  %%CITATION = 00436,4,115;%%

%\cite{Nojiri:2005pu}
\bibitem{Nojiri:2005pu}
  S.~Nojiri and S.~D.~Odintsov,
  %``Unifying phantom inflation with late-time acceleration: Scalar
  %phantom-non-phantom transition model and generalized holographic dark
  %energy,''
  Gen.\ Rel.\ Grav.\  {\bf 38}, 1285 (2006)
  [arXiv:hep-th/0506212].
  %%CITATION = GRGVA,38,1285;%%

%\cite{Capozziello:2005tf}
\bibitem{Capozziello:2005tf}
  S.~Capozziello, S.~Nojiri and S.~D.~Odintsov,
  %``Unified phantom cosmology: inflation, dark energy and dark matter under the
  %same standard,''
  Phys.\ Lett.\  B {\bf 632}, 597 (2006)
  [arXiv:hep-th/0507182].
  %%CITATION = PHLTA,B632,597;%%


%\cite{Nojiri:2006be}
\bibitem{Nojiri:2006be}
  S.~Nojiri and S.~D.~Odintsov,
  %``Modified gravity and its reconstruction from the universe expansion
  %history,''
  J.\ Phys.\ Conf.\ Ser.\  {\bf 66}, 012005 (2007)
  [arXiv:hep-th/0611071].
  %%CITATION = 00462,66,012005;%%


\bibitem{reconstruction}
S.~Nojiri, arXiv:0912.5066. 


%\cite{Capozziello:2010uv}
\bibitem{Capozziello:2010uv}
  S.~Capozziello, J.~Matsumoto, S.~Nojiri and S.~D.~Odintsov,
  %``Dark energy from modified gravity with Lagrange multipliers,''
  arXiv:1004.3691 [hep-th].
  %%CITATION = ARXIV:1004.3691;%%






\end{thebibliography}
\end{document}